\documentclass[fleqn,10pt]{wlscirep}

\usepackage{graphicx}
\usepackage{trfsigns,amsmath}

\title{How to test for partially predictable chaos}

\author[1]{Hendrik Wernecke}
\author[1]{Bulcs\'{u} S\'{a}ndor}
\author[1,*]{Claudius Gros}
\affil[1]{Institute for Theoretical Physics, Goethe University Frankfurt, Germany}
\affil[*]{gros@itp.uni-frankfurt.de}

\begin{abstract}
For a chaotic system pairs of initially close-by trajectories become 
eventually fully uncorrelated on the attracting set. This process of 
decorrelation may split into an initial exponential decrease, characterized by 
the maximal Lyapunov exponent, and a subsequent diffusive process on 
the chaotic attractor causing the final loss of predictability. 
The time scales of both processes can be either of the same or of very 
different orders of magnitude. In the latter case the two trajectories 
linger within a finite but small distance (with respect to the overall 
extent of the attractor) for exceedingly long times and therefore remain 
partially predictable.
\smallskip

Tests for distinguishing chaos from laminar flow widely use the
time evolution of inter-orbital correlations as an indicator. 
Standard tests however yield mostly ambiguous results when it
comes to distinguish partially predictable chaos and laminar flow,
which are characterized respectively by attractors of fractally
broadened braids and limit cycles. For a resolution we introduce 
a novel 0-1 indicator for chaos based on the cross-distance scaling 
of pairs of initially close trajectories, showing that this test 
robustly discriminates chaos, including partially predictable chaos, 
from laminar flow. One can use furthermore the finite time 
cross-correlation of pairs of initially close trajectories to 
distinguish, for a complete classification, also between strong 
and partially predictable chaos. We are thus able to identify 
laminar flow as well as strong and partially predictable chaos in a 0-1 
manner solely from the properties of pairs of trajectories.
\end{abstract}

\begin{document}

\flushbottom
\maketitle
\thispagestyle{empty}



\section*{Introduction}

One characteristic aspect of deterministic 
chaos is the exponential sensitivity of the dynamics to initial 
conditions \cite{tel2006chaotic,alligood1997chaos}. This
sensitivity leads to an effective breakdown of predictability 
as the result of the eventual decorrelation of any given pair 
of trajectories. The decorrelation occurring on a chaotic 
attracting set is measured commonly by the maximal Lyapunov 
exponent\cite{eckmann1992fundamental}, with a positive
value measuring the effective rate of the decorrelation
of initially arbitrary close pairs of trajectories. Other 
standard tests for chaos, such as the correlation 
dimension\cite{grassberger1983characterization} or the 
spectral analysis of the auto-correlation function\cite{alhassid1993onset},
also rely on correlation measures.

Inter-orbital correlations fully decay on chaotic attractors
in the limit of long times. This fact does however not preclude 
the existence of other types of predictable correlations. 
E.\,g.~it is well known\cite{gros2015complex} that 
the sequence $x_n$ produced by the logistic map
$x_{n+1}=rx_n(1-x_n)$ will never decrease twice
in a row for $2<r<4$. In this case we can hence predict 
with $100\%$ confidence that $x_{n+1}$ will be larger 
than $x_n$, if $x_n$ was smaller than $x_{n-1}$, 
even if the system is chaotic. It has also been noted
that correlations may persist for specific chaotic systems
for extended (but finite) times, especially in systems
characterized by multiple time scales \cite{cencini2013finite}
or strong periodic drivings \cite{aizawa1982global}. The
respective cross-correlation $C_{12}$ of initially close 
pairs of trajectories can hence be used as a measure for 
predictability.

In the following we will introduce, distinguish and discuss two types of 
chaotic behavior, denoted partially predictable chaos~(PPC)
and strong chaos respectively, which differ
with respect to what happens for time scales larger 
than the Lyapunov prediction time~$T_\lambda$.

\begin{itemize}
\item {\em Strong chaos}: Predictability vanishes,
      approaching zero on a time scale of $T_\lambda$.
\item {\em PPC}: The first decorrelation occurring on 
      a time scale of $T_\lambda$ does not destroy, in
      this case, all pair-wise correlations. The 
      cross-correlation $C_{12}$ will retain a finite 
      value even for $t\gg T_\lambda$, vanishing 
      only for exceedingly long times.
\end{itemize}

Partial predictability may occur whenever the attracting 
set is characterized by a non-trivial topology. This is 
generically the case for the chaotic state close to a 
period-doubling transition, when the trajectories wander 
chaotically around previously stable limit cycles 
within closed braids\cite{tel2006chaotic}. Our 
observation of partial predictability may hence
be especially of relevance for natural systems
having a tendency to self-organize close to 
criticality\cite{markovic2014power}, as it
implies that the system will hover continuously
close to the transition between laminar and chaotic 
flow\cite{gros2015complex}. Another example of PPC is 
the case of phase locked chaos observed in driven Josephson 
junctions\cite{shukrinov2014structured,kautz1986onset}.

It is generically a challenge to distinguish between partially
predictable chaos and laminar flow on the basis of the maximal 
Lyapunov exponent, which is positive but small for PPC and 
hence difficult to evaluate numerically.
We therefore introduce here a novel test for chaos based on the 
cross-distance scaling of pairs of trajectories. It discriminates
chaos, including PPC and strong chaos, unambiguously in a $0-1$ 
manner. In the following we provide a theoretical motivation for the 
test, a comparison with other measures and an application to the Lorenz 
system\cite{lorenz1963deterministic}, for which we find all 
three types of dynamical regimes: strong chaos, PPC and laminar flow.

Thereafter we combine the indicator for chaos with another 
correlation indicator acting as an effective $0-1$ test, namely 
the finite time cross-correlation of initially close trajectories,
that is able to distinguish strong chaos from PPC. The combination 
of both indicators is capable of drawing an unambiguous distinction 
between all three dynamical phases. As both tests are based on pairs 
of initially close trajectories, requesting only straightforward 
data manipulation, they are easy to implement and suitable both for 
a wide range of scientific fields and for a possible
automation of the procedure.

\begin{figure}[t]\centering
\includegraphics[width=.60\textwidth]{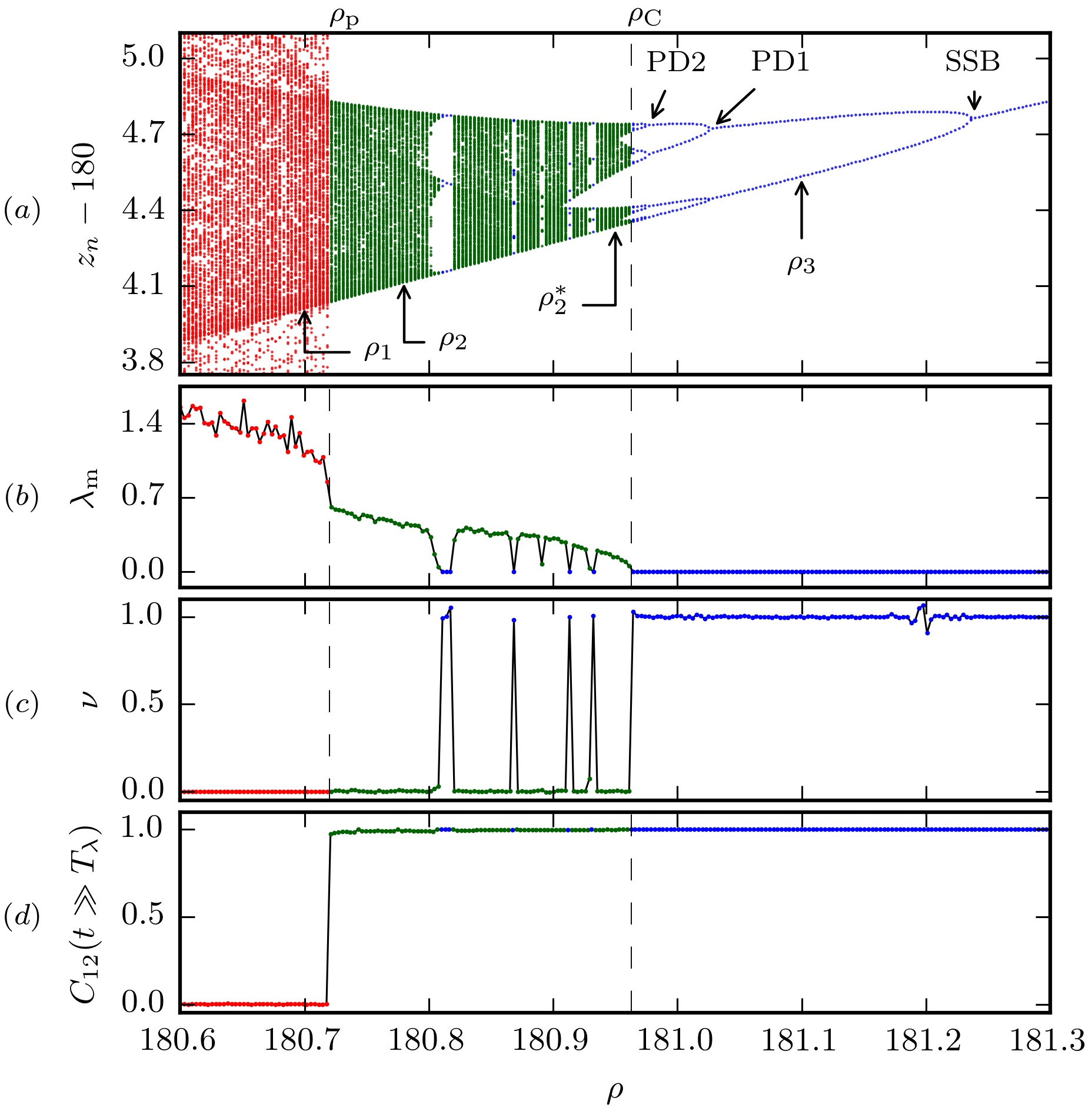}
\caption{\label{fig:lorenz_overview}
Strong chaos (red, for $\rho<\rho_\text{p}\approx180.72$),
partially predictable chaos (green, for $\rho_\text{p}<\rho<\rho_\text{C}\approx 180.96$)
and regular flow (blue, for $\rho_\text{C}<\rho$), for the
Lorenz system (\ref{eq:lorenz}). Shown are (from top to bottom),
$z_n$ from the Poincar\'e section $x=15$ (for a
window of $z_n$), the  maximal Lyapunov exponent~$\lambda_\text{m}$,
the cross-distance scaling exponent~$\nu$, 
see Eq.~(\ref{eq:dist_scaling}), and the cross-correlation $C_{12}(t=200)$,
as defined by Eq.~(\ref{eq:C}).
PD1 and PD2 denote examples of period doubling bifurcations 
and SSB a bifurcation spontaneously breaking the symmetry 
$(x,y,z)\leftrightarrow(-x,-y,z)$ of (\ref{eq:lorenz}).
}
\end{figure}

\section*{Results}

We start by considering with $\mathbf{x}=(x,y,z)$,
\begin{equation}
\label{eq:lorenz} 
\begin{aligned}
\dot{x}&=\sigma(y-x)\,,\quad\qquad
\dot{y}&=x(\rho-z)-y\,,\quad\qquad
\dot{z}&=xy-\beta z
\end{aligned}
\end{equation}
the Lorenz system\cite{lorenz1963deterministic},
which has long been used for studying the
interplay between predictability and 
chaos\cite{palmer1993extended,yadav2005prediction,sparrow2012lorenz}.
We select with $\beta={8}/{3}$ and $\sigma=10$ 
standard parameter settings, retaining
$\rho$ as the bifurcation parameter. 

As an overview we present in Fig.~\ref{fig:lorenz_overview}
the phases of the Lorenz system for a typical parameter 
window $\rho\in[180.6,181.3]$. A transition between two 
types of chaotic regions is observed to occur at 
$\rho_\text{p}\approx 180.72$, together with a transition 
via a cascade of period doubling (halving) bifurcations from chaos 
to laminar flow at $\rho_\text{C}\approx 180.96$.
The dynamics of the intermediate region ($\rho\in[\rho_\text{p}, \rho_\text{C}]$,
green) between strong chaos ($\rho<\rho_\text{p}$, red)
and laminar flow ($\rho>\rho_\text{C}$, blue), is governed by PPC.
A spontaneous symmetry-breaking bifurcation\cite{sandor2015general}~(SSB) is 
additionally shown.
Chaos-chaos transitions involving phase 
space explosions, like the one occurring at $\rho_\text{p}$ 
between partially predictable and strong chaos
(which is in part intermittent\cite{manneville1979intermittency}),
have been studied previously in the context of 
quadratic maps\cite{grebogi1982chaotic} and 
driven Josephson junctions 
\cite{shukrinov2014structured,kautz1986onset}.
They are due to the collision of an unstable manifold 
with the attracting chaotic set, an interior 
crisis\cite{grebogi1982chaotic} which is a
typical example of a global bifurcation\cite{gros2015complex}.
 
\begin{figure*}[t]
\includegraphics[width=.95\textwidth]{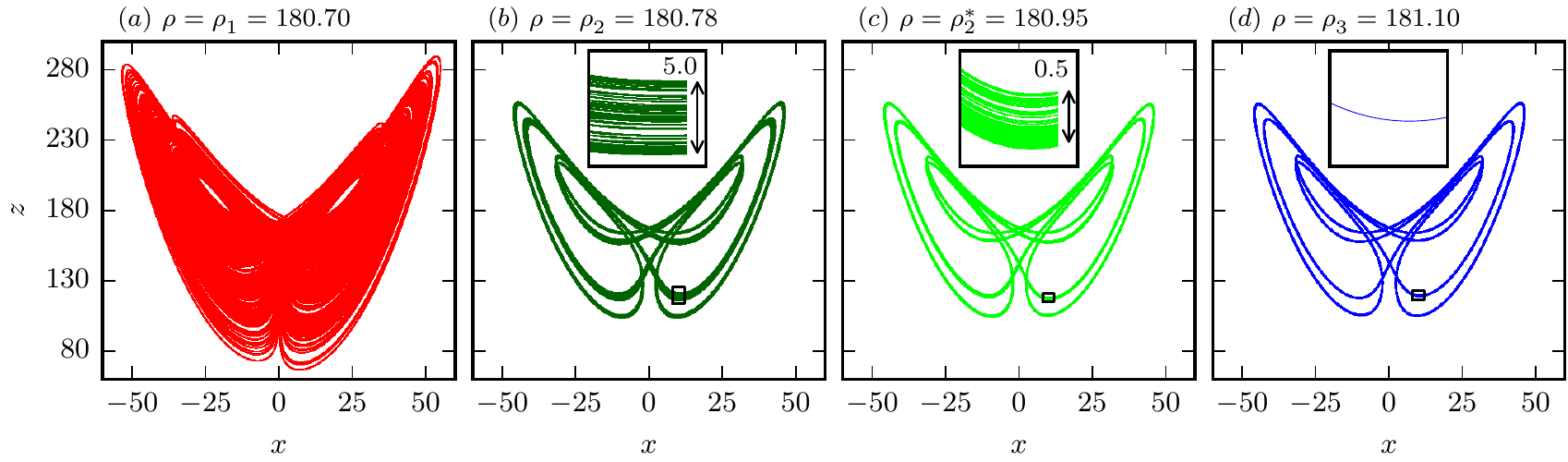}
\caption{\label{fig:lorenz_butterflies}
Sample trajectories of the Lorenz system (\ref{eq:lorenz})
projected to the $x-z$ plane. The phases, from
left to right, $\rho=\rho_1$ (ergodic motion),
$\rho=\rho_2$ and $\rho=\rho_2^*$ (partially predictable chaos) 
and $\rho=\rho_3$ (limit cycle), are indicated in 
Fig.~\ref{fig:lorenz_overview}. 
The arrows in the respective insets indicate the width of
the (fractal) braids in the partially predictable phase. 
The respective standard deviations $s$ of the
attractors are $s\in[58.76,58.92]$ for 
$\rho\in[\rho_1,\rho_3]$.
}
\end{figure*}

In Fig.~\ref{fig:lorenz_butterflies} we present the projections 
to the $x-z$ plane of the respective attracting sets for 
($a$) strong chaos, ($b$) and ($c$) PPC, and ($d$) laminar 
flow. Partially predictable chaos is at times difficult
to distinguish visually from laminar flow (compare 
Figs.~\ref{fig:lorenz_butterflies}~($c$) and ($d$)),
we hence provide the respective blow-ups in the insets.
The partially predictable chaotic attractors can be thought
as fractally broadened limit cycles, viz as braids.

\begin{table}[b]\centering
\caption{
The average maximal Lyapunov exponent $\lambda_\text{m}$ together
with the second- and third largest Lyapunov exponent,
$\lambda_2$ and $\lambda_3$ (compare Fig.~\ref{fig:lyapspec}).
The respective Lyapunov prediction times $T_\lambda$ have been 
evaluated for the two sets of initial distances~$\delta$ used
in Fig.~\ref{fig:distanceScaling}. The values of $\rho$
used are indicated in Fig.~\ref{fig:lorenz_overview}.
\label{tab:lyapunov}
}
\begin{tabular}{rccccc}
\toprule
 &$\lambda_\text{m}$
 &$\lambda_\text{2}$
 &$\lambda_\text{3}$
&\multicolumn{2}{c}{$T_\lambda(\delta)$}\\
 &&&&$\delta=10^{-5}$&$\delta=10^{-8}$\\\hline
 $\rho_1=180.70$   & 1.18 & \phantom{-}0.00 & -14.85 &$\phantom{0}3.9$&$\phantom{0}9.8$\\
 $\rho_2=180.78$   & 0.47 & \phantom{-}0.00 & -14.14 &$\phantom{0}9.8$&$24.5$\\
 $\rho_2^*=180.95$ & 0.14 & \phantom{-}0.00 & -13.81   &$32.9$&$82.2$\\
 $\rho_3=181.10$   & 0.00 &           -0.60 & -13.07   &-&-\\
\bottomrule
\end{tabular}
\end{table}

The maximal Lyapunov exponent\cite{klages2010deterministic}~$\lambda_\text{m}$
presented in 
Fig.~\ref{fig:lorenz_overview}~($b$)
has been evaluated by extracting the initial slope of 
the logarithmic distance 
$\langle\ln(|\mathbf{x}_1(t)-\mathbf{x}_2(t)|)\rangle$
of two trajectories, as averaged over $10^4$ pairs 
with initial distances of $\delta=10^{-8}$, when
plotted  as a function of time. $\lvert\ldots\rvert$
denotes here the Euclidean distance and $\langle\ldots\rangle$ 
the average over initial conditions on the attractor sampled 
with the natural distribution\cite{tel2006chaotic}
(the natural invariant measure\cite{hunt2013theory}).
We used in addition $10^4$ pairs of trajectories for the 
cross-correlation $C_{12}(t=200)$, see Eq.~(\ref{eq:C}). 
The choice of $t=200$ has been made in order to ensure that 
we neither have to deal with initial effects nor with 
numerical inaccuracies, the latter due
to the chaotic nature of the flow.
 
We have also evaluated a scaling exponent $\nu$ (discussed 
further below, see Eq.~(\ref{eq:dist_scaling})), which characterizes
the scaling of the long-term distance between two trajectories. 
Our choice to favor the average logarithmic distance for computing
the maximal Lyapunov exponent over more sophisticated methods 
is motivated by a conceptual computational aspect: in this way 
all three indicators presented here, i.\,e.\ the maximal 
Lyapunov exponent~$\lambda_\text{m}$, the cross-correlation $C_{12}$
and the cross-distance scaling exponent $\nu$, can be evaluated
from the time evolution of initially close-by trajectories.
In the Methods section we compare this approach for computing the 
maximal Lyapunov exponent to the results obtained by
Benettin's method\cite{benettin1980lyapunov,skokos2010lyapunov}.

Two fundamental time scales determine the initial dynamics.
The first is the quasi-period $\tau$, which is the average
time a trajectory needs to come back to the same intersection of 
the braid with the Poincar\'{e} plane. It is comparable to the period 
of the limit cycle and we find $\tau\simeq2.2$ to hold for all 
partially predictable attractors $\rho_\text{C}<\rho<\rho_\text{p}$.
The second time scale is the Lyapunov prediction\cite{karolyi2010finite}
time $T_\lambda=
\ln\left(\lvert\mathbf{x}_1-\mathbf{x}_2\rvert/\delta\right)/\lambda_\text{m}$,
which is the time it takes for two exponentially diverging trajectories
starting from an initial separation $\delta$ to reach a given 
finite distance $\lvert\mathbf{x}_1-\mathbf{x}_2\rvert$. For 
these two distances we used $\delta=10^{-8}$ 
and $\lvert\mathbf{x}_1-\mathbf{x}_2\rvert\sim 0.001$
respectively. At the latter distance a finite amount of 
predictability is lost, viz the cross-correlation $C_{12}$,
as defined by Eq.~(\ref{eq:C}), starts to deviate 
from unity. Given the values of the maximal Lyapunov 
exponent~$\lambda_\text{m}$ presented in 
Fig.~\ref{fig:lorenz_overview} we obtain 
$T_\lambda\approx10$ and
$T_\lambda\approx25$ for 
$\rho=\rho_1=180.70$ and $\rho=\rho_2=180.78$ 
respectively (cf.\ also Table~\ref{tab:lyapunov}). The initial loss of
predictability occurs 
hence after a few cycles around the braid.


\subsection*{Cross-distance scaling}
\begin{figure}[t]
\includegraphics[width=0.99\textwidth]{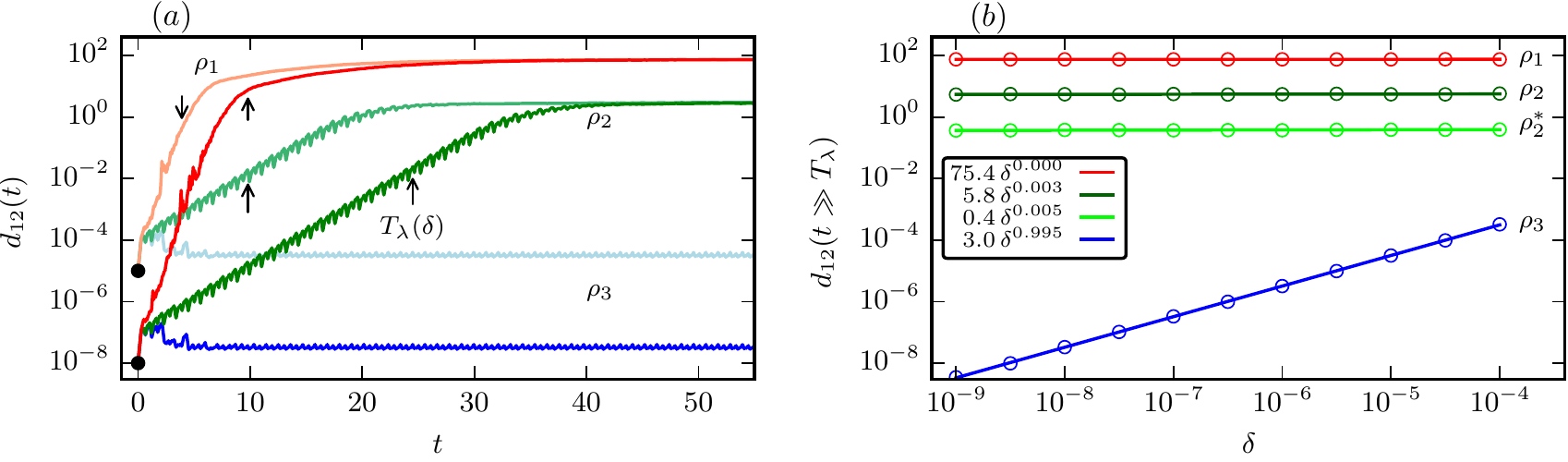}
\caption{\label{fig:distanceScaling}
Distance scaling of initially close-by trajectories.
($a$) The distance $d_{12}(t)$ of initially close-by pairs of 
trajectories, averaged over $10^{4}$ initial
conditions, with initial distances 
$d_{12}(0)=\delta=10^{-8}$ and $\delta=10^{-5}$ (black bullets). 
For the regular flows ($\rho=\rho_3=181$, blue lines) the 
long-term distance depends on $\delta$. The strongly chaotic 
attractor ($\rho=\rho_1=180.7$, red lines) approaches the maximal 
distance independently of the initial distance within the respective
Lyapunov prediction time $T_\lambda$. In the case of PPC 
($\rho=\rho_2=180.78$, green lines) the distances reach a
quasi stationary plateau that is independent of the initial 
distance. For comparison we marked the Lyapunov prediction 
times $T_\lambda$ at the respective curves by arrows.
($b$)~The scaling behavior, see Eq.~(\ref{eq:dist_scaling}), 
of the averaged long-term distance $d_{12}(t=200)$.
The results (circles) are for the strongly chaotic phase 
($\rho=\rho_1$, top), for the partially predictable chaos 
($\rho=\rho_2$ and $\rho=\rho_2^*$, middle) and for 
a limit cycle ($\rho=\rho_3=181.10$, bottom), as indicated in 
Fig.~\ref{fig:lorenz_overview}. The respective solid lines 
are linear fits to the log-log plot.
}
\end{figure}

A large body of work\cite{grassberger1983characterization,eckmann1992fundamental}
has shown that strange attractors are relatively difficult to characterize 
in detail, even for low-dimensional dynamical systems.
For systems with a higher dimension, such as autonomous
neural networks or climate models, it may even
be a challenge to robustly distinguish laminar from 
chaotic flows. Here we propose that the scaling of the 
long-term distance $d_{12}(t\gg T_\lambda)$
of two trajectories,
\begin{equation}
\label{eq:dist_scaling}
d_{12}(t\gg T_\lambda) \propto \delta^\nu,
\quad\qquad
d_{12}(t)=\langle\lvert\mathbf{x}_1(t)-\mathbf{x}_2(t)\rvert\rangle\,,
\end{equation}
may be used as a reliable indicator for chaos,
where we denote with $d_{12}(t=0)=\delta$ the initial distance,
and with $\nu$ the cross-distance scaling exponent.

For an illustration of how the long-term distance 
$d_{12}(t\gg T_\lambda)$ depends on the initial 
distance $\delta$, we show in Fig.~\ref{fig:distanceScaling}~($a$) 
the time evolution of the distance~$d_{12}$ between pairs 
of trajectories, considering initial distances 
$\delta=10^{-8}$ and $\delta=10^{-5}$, as averaged over 
$10^{4}$ pairs. For strong chaos (red curves, $\rho=\rho_1$)
and PPC (green curves, $\rho=\rho_2$) the long-term distance
does not depend on the initial distance. The scaling exponent 
thus vanishes, $\nu=0$, for chaotic motion. The initial 
slope of the curves reflects the exponential divergence 
of chaotic trajectories within the time scale of the Lyapunov 
prediction time $T_\lambda$. For the laminar flow (blue curves, 
$\rho=\rho_3$) the long-term distance depends on the other side 
on $\delta$, leading to a non-zero scaling exponent, $\nu\neq0$.

For the results shown in Fig.~\ref{fig:distanceScaling}~($b$)
we have evaluated for every $\rho$ considered
the long-term distance~$d_{12}(t=200)$ starting from initial
distances $\delta\in[10^{-9},10^{-4}]$, averaging
each time over $10^{3}$ pairs of trajectories.
We note that the scaling exponent~$\nu$ can be
extracted reliably from a linear regression of 
the data in a log-log plot when the initial distance~$\delta$
is smaller than the distance of two neighboring attractors
or parts of the same attractor.
Additionally we note that the choice of $t=200$ 
was selected such that $t\gg T_\lambda$ holds for the interval
of $\rho$ considered (cf.\ Table~\ref{tab:lyapunov}).
Close to the period doubling transition to chaos, viz for
$\rho\lesssim\rho_\text{C}$, the maximal Lyapunov exponent 
becomes very small ($\lambda_\text{m}\lesssim10^{-2}$) and
the Lyapunov prediction times large ($T_\lambda>200$). In this  
case a larger time $t\gg T_\lambda$ would be needed.

The linear scaling $\nu=1$ observed for the limit cycle 
($\rho=\rho_3$) in Fig.~\ref{fig:distanceScaling}~($b$)
stems from the fact that any two orbits attracted 
by a limit cycle follow each other perpetually, with 
the average final separation being proportional to the 
initial separation. This relation can be motivated 
analytically using a local approximation to the attracting 
set in the normal form of limit cycles 
(cf.\ the Methods section). 

For chaotic phases the long-term average distance settles 
on the other hand to a finite value determined by the extent 
of the attracting set, independently of the initial distance~$\delta$, 
leading to a vanishing scaling exponent~$\nu=0$.
As observed in Fig.~\ref{fig:distanceScaling}~($a$)
the time needed for strong chaos to reach long-term stationarity in 
$d_{12}(t)$ is proportional to the Lyapunov prediction time $T_\lambda$.
For PPC the long-term limit is however only
reached for $t>T_\text{PPC}$, where the decorrelation 
time $T_\text{PPC}$, i.\,e.\ the time that a pair of trajectories
needs to get fully uncorrelated, is significantly longer than
both the quasi-period $\tau$ and the Lyapunov prediction time 
$T_\lambda$. The scaling exponent~$\nu\approx0$ is hardly 
distinguishable from zero for measurements at time 
$t\in[T_\lambda,T_\text{PPC}]$. We remark that $d_{12}(t\to\infty)$ 
is however determined by the overall 
extent of the attracting
set in the limit of large times. For 
times $t>T_\lambda$, right after the initial exponential 
decorrelation, the typical separation of two orbits
$d_{12}(t)$ is of the order of the braid width
(cf.\ insets in Fig.~\ref{fig:lorenz_butterflies} ($b$), ($c$)).

In Fig.~\ref{fig:lorenz_overview}~($c$)
the cross-distance scaling exponent $\nu$
for the entire range of $\rho$ considered here
is shown. We note, that the transition from
chaos to laminar flow occurring at $\rho_\text{C}\approx 180.96$
is accompanied by a sudden jump in $\nu$
from zero to one. This is quite remarkable, as
the corresponding maximal Lyapunov exponent
$\lambda_\text{m}$, also shown in Fig.~\ref{fig:lorenz_overview},
becomes, on the other hand, continuously smaller
when approaching $\rho_\text{C}$ from the chaotic side.

The cross-distance scaling is a robust $0-1$~test for chaos 
that also classifies PPC correctly 
\cite{barrio2009spurious}.
For a further evaluation we applied it to the chaotic 
states found in previously studied neural networks 
\cite{wernecke2016attractor}, which we generalized
in size (with up to 300 dimensional phase spaces).
We also examined the three-dimensional Shilnikov attractor\cite{guan2013non} 
(cf.\ Appendix~\ref{appsec:altSys}), as it is similar to the Lorenz 
system, albeit with all degrees of freedom evolving on the same time scale.
For both systems the test presented here worked without problems.

For a comparison we applied the Gottwald 
$0-1$~test\cite{gottwald2004new,gottwald2009implementation}
to the three different dynamical regimes of the Lorenz system
presented above (cf.\ Fig.~\ref{fig:lorenz_butterflies}).
Using Gottwald's method we were able to classify regular 
motion $\rho=\rho_3$ and strong chaos $\rho=\rho_1$ correctly,
but not partially predictable chaos. For $\rho=\rho_2$ even
an exceedingly long run time, $t=10^6$, did not provide a clear result.

\subsection*{Cross-correlation of initially close trajectories}

\begin{figure*}[t]
\includegraphics[width=.95\textwidth]{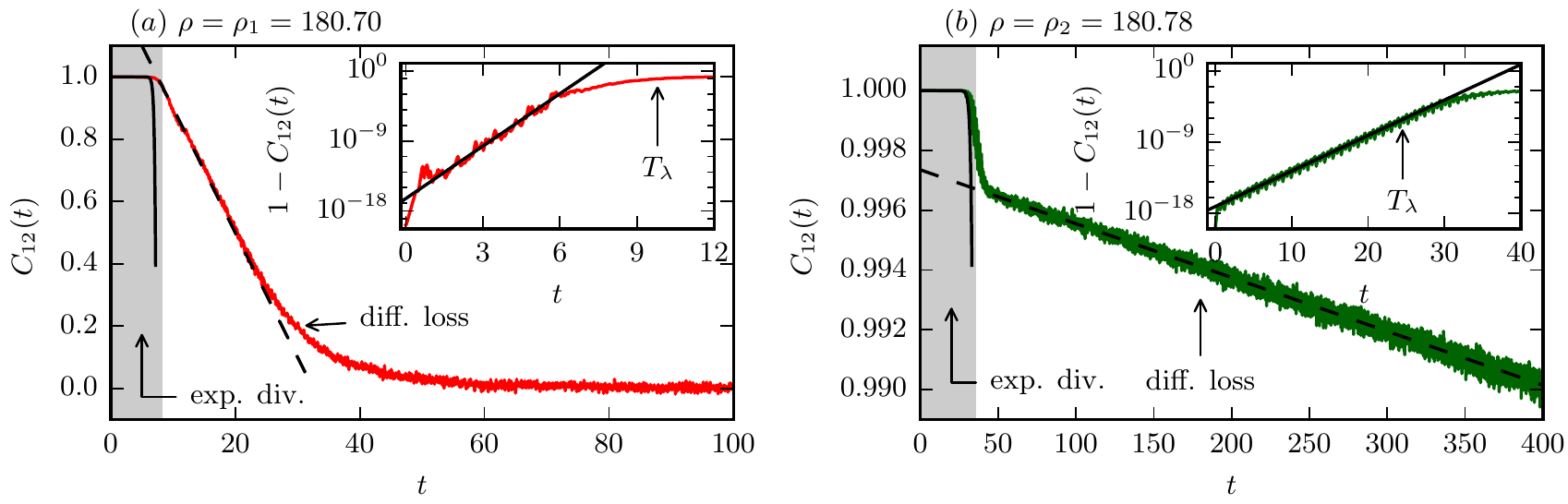}
\caption{\label{fig:C12_time} 
The cross-correlation $C_{12}$ for pairs of trajectories 
with an initial distance $\delta=10^{-8}$ and
averaged over $10^4$ pairs over time. The data is for ($a$)~strong chaos 
($\rho=\rho_1=180.70$) and ($b$)~partially 
predictable chaos ($\rho=\rho_2=180.78$);
compare Fig.~\ref{fig:lorenz_overview}. For both cases we 
find an initial time interval of exponential decorrelation
(shaded gray).
The blow-ups magnifying that region show that the correlation
$C_{12}\approx1$ initially decreases
with $1-2.8\cdot10^{-13}\,e^{\lambda_\text{C}t}$, $\lambda_\text{C}=5.17$ 
(inset in ($a$)) and as 
$1-4.7\cdot10^{-14} e^{\lambda_\text{C}t}$, $\lambda_\text{C}=1.04$ 
(inset in ($b$)), as indicated by the respective solid lines.
The linear decrease of $C_{12}$ for intermediate times
corresponding to a diffusive loss of predictability is
$\propto(-0.04t)$ in ($a$) and $\propto(-1.8\cdot10^{-5}t)$
in ($b$), as indicated by the respective dashed lines.
}
\end{figure*}

An important point for real-world applications
are the long-term repercussions of variations
in the initial conditions. For concreteness 
we consider with
\begin{equation}
C_{12}(t) = \left\langle
      (\mathbf{x}_1(t)-\boldsymbol{\mu})\cdot
      (\mathbf{x}_2(t)-\boldsymbol{\mu})
\right\rangle/s^2\,,
\label{eq:C}
\end{equation}
the cross-correlation function of two bounded
and initially close-by trajectories $\mathbf{x}_1(t)$ 
and $\mathbf{x}_2(t)$. Here $\langle\ldots\rangle$ 
denotes an average over initial conditions on 
the attractor sampled with the natural distribution,
$\boldsymbol{\mu}$ the center of gravity, and 
$s$ the average extent
of the attracting set,
\begin{equation}
\boldsymbol{\mu}=\lim_{T\to\infty}\frac{1}{T}
\int_T^{2T}\!\!\!\mathbf{x}(t)\,\mathrm{d}t,\quad\qquad
s^2=\lim_{T\to\infty}\frac{1}{T}
\int_T^{2T}\!\!\!\left[\mathbf{x}(t)-\boldsymbol{\mu}\right]^2\,\mathrm{d}t\,.
\label{eq:musig}
\end{equation}
The cross-correlation is normalized to unity for
close-by trajectories, i.\,e.\ for
$|\mathbf{x}_1(t)- \mathbf{x}_2(t)|\to0$.
For chaotic attracting sets the cross-correlation 
$C_{12}$ vanishes in the long-term limit 
$t\to\infty$, with a finite $C_{12}\ne0$ implying
finite amounts of predictability.

For a geometric comparison we define the averaged square 
distance 
$D_{12}(t)=\langle[\mathbf{x}_1(t)-\mathbf{x}_2(t)]^2\rangle$
between two trajectories, which leads, when using 
(\ref{eq:C}), to
\begin{equation}
D_{12}(t)=2s^2[1-C_{12}(t)]\;.
\label{eq:D_12}
\end{equation}
For large cross-correlations $C_{12}\to1$ the two trajectories
are close-by with respect to the overall
extent~$s$ of the
attracting region, in the sense that $D_{12}\ll s^2$.

It  is evident from Fig.~\ref{fig:lorenz_butterflies}, that 
the overall shape of the attractor changes little across
the transition from laminar flow ($\rho=\rho_3$) to chaos 
($\rho=\rho_2^*$), and that the previously one dimensional
attracting state (the limit cycles) does broaden to a 
closed chaotic braid. This behavior can also be 
viewed as chaotic wandering around limit 
cycles\cite{tel2006chaotic}.

In Fig.~\ref{fig:C12_time} we present the time evolution
of the cross-correlation $C_{12}$ for the case of strong chaos,
$\rho=\rho_1$ in ($a$), and partially predictable chaos, $\rho=\rho_2$
in ($b$). We note that $C_{12}$ remains close to full predictability, 
$C_{12}\simeq1$, within the respective time scales of the Lyapunov 
prediction time~$T_\lambda$. We have included in both panels of 
Fig.~\ref{fig:C12_time} fits to the cross-correlations of the 
form $1-c\exp(\lambda_\text{C}t)$, an approximation resulting 
from (\ref{eq:D_12}), where 
$\lambda_\text{C}\geq2\lambda_\text{m}$ \cite{badii1988correlation}
($c$ is a fit parameter).  $C_{12}$ decreases linearly 
for times larger than $T_\lambda$, saturating eventually 
to zero when full decorrelation is achieved 
\cite{crisanti1991lagrangian,boffetta2004introduction}:
\begin{equation}
D_{12}(t)/2s^2=1-C_{12}(t)\propto\left\{
\begin{array}{cl}
\operatorname{e}^{\lambda_\text{C}t}&\text{ for }t<T_\lambda\\
t&\text{ for }t>T_\lambda
\end{array}\right.\,.
\end{equation}
For strong chaos both the exponential divergence and 
the diffusive loss of predictability happen on the same 
time scale. We remark here that the evolution of the
cross-correlation presented in Fig.~\ref{fig:C12_time} 
for the case of strong chaos bears a surprising similarity 
with the measured relative accuracy of weather forecasting 
over a period of two weeks\cite{gros2012pushing,simmons2002some}.

For partially predictable attractors, $\rho=\rho_2$, we find
qualitatively the same behavior as for strong chaos, at least
as matter of principle, with a dramatic separation of time
scales setting however in beyond the initial phase of exponential 
divergence. The slope of the linear decrease is, as evident from 
Fig.~\ref{fig:C12_time}~($b$), three orders of magnitude 
smaller in the partially predictable case ($1.8\cdot10^{-5}$ 
instead of $4\cdot10^{-2}$).

PPC can be found also in systems controlled by a single
microscopic time scale (cf.\ Appendix~\ref{appsec:altSys}). 
We hence attribute the emergence of PPC to the fact, that the 
attractor is topologically equivalent, for $\rho=\rho_2$, to 
elongated closed braids.
Comparing the braid width from the insets in 
Fig.~\ref{fig:lorenz_butterflies} to the linear 
distance $d_{12}(t=200)$ in Fig.~\ref{fig:distanceScaling}~($b$)
($5$ and $0.5$ in comparison to $5.8$ and $0.4$ for
$\rho=\rho_2$ and $\rho=\rho_2^*$ respectively), we find 
that the initial exponential divergence occurs dominantly 
perpendicular to the braid. Once the separation of two 
trajectories has reached the braid width it can increase 
further only along the braid, which is in turn a diffusive process
and hence slow. This means that the chaotic flow remains
partially predictable for remarkable long times compared to the 
Lyapunov prediction time $T_\lambda$. From the linear fit in 
Fig.~\ref{fig:C12_time}~($b$) we estimate that it takes
$T_\text{PPC}\approx10^{4}$ until correlations vanish
effectively for the partially predictable case $\rho=\rho_2$.

The cross-correlation $C_{12}(t=200)$ shown in the
Fig.~\ref{fig:lorenz_overview}~($d$) vanishes 
for $\rho<\rho_p$, which is hence a phase in which
predictability is lost for times larger
than the Lyapunov prediction time $T_\lambda$.
The ergodicity of pairs of trajectories is however broken for 
intermediate times $T_\lambda < t < T_\text{PPC}$
in the PPC phase realized for $\rho_p<\rho<\rho_\text{C}$.
Partially predictable chaos is hence characterized both 
by positive Lyapunov exponents $\lambda_\text{m}>0$ and 
by a finite predictability coefficient 
$C_{12}(t\gg T_\lambda)\ne0$. This notion of
predictability does naturally not exclude the
possibility of finding additional finite time 
windows of predictability due to the presence of
periods of quasi-laminar flow embedded in 
the overall chaotic time evolution\cite{palmer1993extended}.
Measuring the fractal dimension \cite{eckmann1992fundamental}
with the box-counting method we find that the attractors 
in the PPC phase have fractal dimensions slightly larger
than two, as usual for the Lorenz system\cite{mcguinness1983fractal}.

The exceedingly slow loss of predictability occurring for
$\rho=\rho_2$ can be observed also in systems in which
all defining dynamical parameters are of the same order
of magnitude (cf.\ Appendix~\ref{appsec:altSys}). The magnitude 
of the respective diffusion coefficient is hence only indirectly 
related, for the case of the Lorenz system, to the relative size 
of $\beta$, $\sigma$ and $\rho$ in (\ref{eq:lorenz}).
We also note that the neutral flow along the braids,
i.\,e.\ the flow along the attractor which is characterized by 
a vanishing average $\lambda_2=\langle\lambda^{(\text{l})}_2\rangle=0$ 
of the second-largest local Lyapunov exponent $\lambda^{(\text{l})}_2$, 
is highly dispersive (cf.\ Fig.~\ref{fig:lyapspec} in the Methods section).
Additionally we remark that PPC manifests itself in a linear
decrease of the amplitudes of the periodic oscillation of the 
auto-correlation function (cf.\ Methods section). 


\section*{Discussion}

We have proposed here a new $0-1$~test for
the occurrence of chaos derived from the long-term 
scaling behavior of the distance between pairs of
initially close trajectories. We find the $0-1$ test 
to be extraordinarily robust and that chaotic dynamics
may be partially predictable whenever two initially close 
trajectories remain within a finite but small distance for 
extended periods. Partial predictability occurs when 
the initial exponential divergence stops at length scales
which are finite but substantially smaller than the overall
extent of the attracting set. For the Lorenz system we found 
that residual predictability levels of the order of 99\% 
are retained despite non-zero Lyapunov exponents 
\cite{klages2013weak} and that the system is stable in
this state against finite perturbations 
\cite{boffetta2002predictability,cencini2013finite}.
The notion of partial predictability implies
macroscopic predictability in terms of coarse 
grained predictions. Taking the case of weather forecasting, 
which is plagued notoriously by chaotic instabilities
\cite{palmer1993extended,shukla1998predictability,gros2012pushing},
it may hence be possible to predict with confidence the formation of
a low pressure area, to give an example, but not its exact 
extension and depth.

We have shown here that partial predictability 
is not a consequence of varying local Lyapunov 
exponents on the attracting set and that the averaged 
Lyapunov exponents in terms of the Lyapunov spectrum 
yield prediction times which are orders of magnitude
smaller than the time scales observed for partial
predictability. Partial predictability is essentially 
a consequence of topological constraints, e.\,g.\ when
chaotic braids arise from a previous period doubling transition.
PPC is hence expected to be found for a wide range of systems, 
such as enzyme reactions\cite{geest1992period} and models 
of asset pricing \cite{brock1998heterogeneous}. In this context we point out
that indications for partially predictable chaos 
have been found recently in the phase space of the sensorimotor
loop of simulated self-organized robots \cite{martin2016closed}.
It would be interesting to investigate in further studies whether 
the concept of partial predictability, which does not require 
multiple time scales {\it per se}, could be generalized 
to time dependent snapshot 
\cite{romeiras1990multifractal,bodai2012annual} or pullback 
attractors \cite{ghil2008climate} arising in stochastic and/or
driven chaotic systems.

The dynamical regimes discussed here -- strong 
chaos, PPC and laminar flow -- can be distinguished
when combining the $0-1$ test for chaos with an analysis of the 
long-term saturation plateau of the inter-trajectory 
cross-correlation function $C_{12}(t)$, which may be 
finite (for PPC and laminar flow) or zero (for strong 
chaos). We also stress that the three indicators -- 
global maximal Lyapunov exponent, cross-correlation
and cross-distance scaling -- examined in this work
rely on the evolution of initially close pairs of 
trajectories. These indicators can hence  be evaluated
by a straightforward manipulation of the data without
the need to investigate further the nature of the attracting 
set. It is possible to automatize the computation of there
examined $0-1$ indicators -- we provide a pseudo
code routine in Appendix~\ref{appsec:automat} -- and
to obtain thus a combined test capable to distinguish
the three dynamical regimes characterizing
dissipative autonomous dynamical systems.


\section*{Methods}

All computations that involved solving Eq.~(\ref{eq:lorenz})
were performed using a Runge-Kutta-Fehlberg 
algorithm\cite{fehlberg1969low} of order 4/5 and step 
size $\varDelta t=10^{-3}$. Testing the accuracy of the 
results by systematically varying $\varDelta t$ we
found that the limitations due to the chaotic nature of
the motion allow for reliable results for integration 
times up to $t\sim500$.

\subsection*{Derivation of the cross-distance scaling for limit cycles}
\label{appsec:scaling}

Above we showed that the long-term distance
$d_\infty = \lim\limits_{t\to\infty}d_{12}(t)$ for 
of two initially close-by trajectories scales linearly with 
the initial distance~$\delta$ whenever the dynamics settles 
in an attracting limit cycle. For an analytic understanding 
of this observation we consider the two dimensional normal form
for limit cycles in polar coordinates $(\varphi,r)$\cite{gros2015complex},
\begin{equation}
\label{eq:psys}
\dot{\varphi}=\Omega(r)\,,\quad\qquad
\dot{r}=r(\Gamma^2-r^2)\,,
\end{equation}
where the time evolution of the angle~$\varphi$ is described by an arbitrary
smooth function $\Omega(r)$ of the radius $r$,

\begin{figure*}[t]
\includegraphics[width=.99\textwidth]{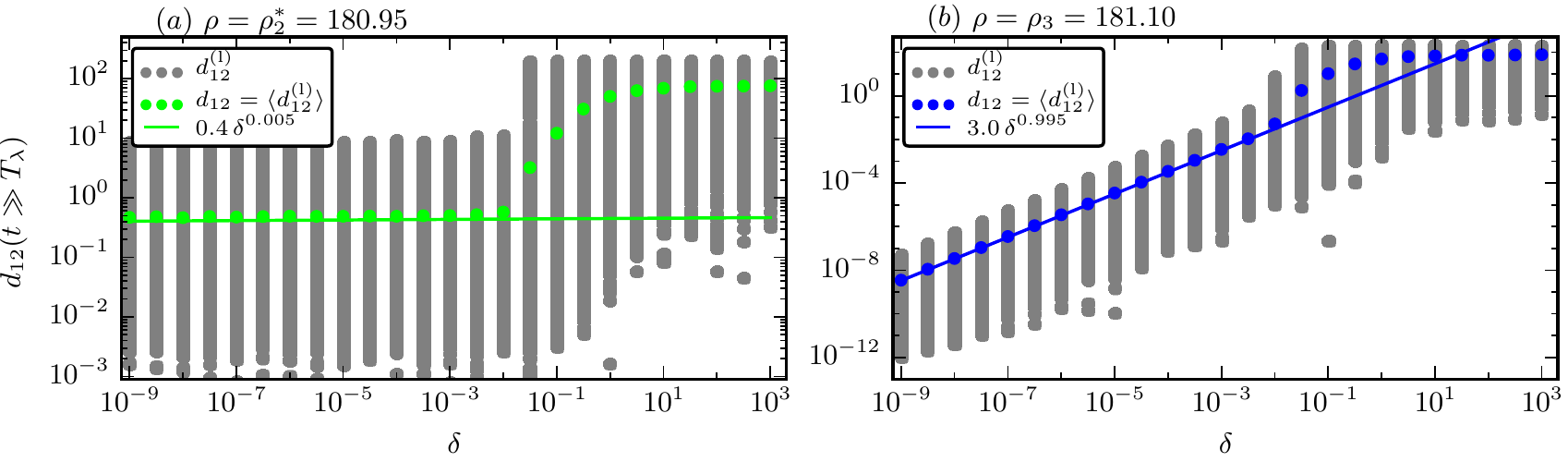}
\caption{\label{fig:breakdown}
The scaling behavior of the averaged long-term distance $d_{12}(t=200)$ 
(colored circles), see Eq.~(\ref{eq:dist_scaling}), with the 
respective fits on a log-log scale (fitted for $\delta<10^{-2}$)
(solid lines, cf.~Fig.~\ref{fig:distanceScaling}). The distribution 
of the non-averaged distances~$d_{12}^{(\mathrm{l})}(t=200)$,
as obtained from $1000$ initial conditions, are shown in
addition (gray circles).
($a$) For PPC with $\rho=\rho_2^*=180.95$ one finds, in agreement with 
Fig.~\ref{fig:distanceScaling}, a close to constant cross-distance 
scaling.
($b$) For a laminar flow with $\rho=\rho_3=181.10$ the scaling exponent 
is $\nu\approx1$ for distances~$\delta<10^{-2}$.
In both cases the scaling breaks down when a symmetry related close-by 
attractor starts to attract a fraction of the orbits for $\delta>10^{-2}$.
}
\end{figure*}

Expanding Eq.~(\ref{eq:psys}) to first order around the
limit cycle $r(t)\equiv\Gamma$, viz using $r=\Gamma+\epsilon$,
we find
\begin{equation}
\label{eq:appsys}
\dot{\varphi}=\Omega(\Gamma)+\Omega^\prime(\Gamma)\,\epsilon\,,\quad\qquad
\dot{\epsilon}=-2\Gamma^2\,\epsilon
\end{equation}
for the behavior in the close neighborhood of the limit cycle,
with $\Omega^\prime=\mathrm d\Omega/\mathrm dr$ denoting the derivative 
with respect to the radius~$r$. Substituting $(\varphi,\epsilon)\rightarrow(x,y)$
and $(\Omega,\Omega',2\Gamma^2)\rightarrow(a,b,c)$
we then obtain with
\begin{equation}
\label{eq:modsys}
\dot{x}=a+b\,y\,,\quad\qquad
\dot{y}=-c\,y
\end{equation}
the Cartesian normal form of a limit cycle. The parameter $a$ 
hence represents the base speed of the flow along the limit cycle, $b$
the rate with which the flow parallel to the limit cycle changes
with the distance $y$ from the limit cycle, and $c$ the time scale
needed to relax to the attractor. The solution $(x(t), y(t))$ of 
the linearized system with the initial conditions 
$(x,y)(t=0)=(x_\text{o},y_\text{o})$ is given by
\begin{equation}
\label{eq:sol}
x(t)=x_\text{o}+at+\frac{b}{c}y_\text{o}\left(1-\e^{-ct}\right)\,,\quad\qquad
y(t)=y_\text{o}\,\e^{-ct}\,.
\end{equation}
As we are interested in the behavior of two initially close trajectories
(of which both are close to the attractor), we consider two trajectories starting
from $(x_\text{o},y_\text{o})$ and $(x_\text{o}+\delta_x,y_\text{o}+\delta_y)$.
Here $\delta_x$ and $\delta_y$ denote the initial distances 
between the trajectories in their respective dimension and 
$\delta=(\delta_x^2+\delta_y^2)^{1/2}$ the initial 
Euclidean distance between the trajectories. Both trajectories converge
in the long-term limit $t\to\infty$ to the limit cycle $y\to0$.
The Euclidean distance between the trajectories is hence given by
\begin{equation}
d_{12}(t)=\left([x_1(t)-x_2(t)]^2+[y_1(t)-y_2(t)]^2\right)^{1/2}
=\left(\left[\delta_x+\frac{b}{c}\delta_y(1-\e^{-ct})\right]^2+
\delta_y^2\e^{-2ct}\right)^{1/2}\label{eq:dist1}\,.
\end{equation}
The distance approaches a finite value in the long-term 
limit $t\to\infty$, which we term the long-term distance
\begin{equation}
d_\infty(\delta_x, \delta_y)=\lim\limits_{t\rightarrow\infty}d_{12}(t)
=\left|\delta_x+\frac{b}{c}\delta_y\right|\label{eq:ltdist}\,,
\end{equation}
where $\lvert\cdot\rvert$ denotes the modulus.

Averaging $d_\infty(\delta_x, \delta_y)$ over a 
circle $\mathcal{C}$ centered around $(x_\text{o},y_\text{o})$,
defined by $\delta^2=\delta_x^2+\delta_y^2$,
we obtain
\begin{equation}
\left<d_\infty\right>=\frac{1}{2\pi\delta}
\oint\limits_\mathcal{C}\mathrm ds\;d_\infty(\delta_x,\delta_y)
=\frac{2}{\pi}\left(\frac{b^2}{c^2}+1\right)^{1/2}\,\delta
\label{eq:avdistres}\,.
\end{equation}
The average long-term distance~$\left<d_\infty\right>$ is hence proportional
to the initial distance~$\delta$, with the constant of proportionality 
$2(b^2/c^2+1)^{1/2}/\pi$ depending through $b/c$ on the properties of the 
flow close to the limit cycle. The factor $b/c$ can be smaller 
or larger than unity, implying that the long-term distance 
of the two trajectories may exceed the initial distance.

The normal form (\ref{eq:modsys}) describes the local flow close
to a limit cycle. For the case of a non-uniform base velocity 
$a=a(x)$ one need to generalize (\ref{eq:avdistres})
by averaging over full periods.

\subsection*{Choice of initial distances}

The cross-distance scaling (cf.~Eq.~(\ref{eq:dist_scaling})) is valid
only when the two trajectories considered are attracted by
the same attractor. This condition is satisfied for the
values of $\delta$ considered in Figs~\ref{fig:lorenz_overview} 
and \ref{fig:distanceScaling}, namely
$\delta\in[10^{-9},10^{-4}]$, but not necessarily for
larger values of $\delta$, as illustrated in
Fig.~\ref{fig:breakdown}.

For the partially predictable chaotic attractor, with $\rho=\rho_2^*=180.95$ in
Fig.~\ref{fig:breakdown}~($a$), we find the expected scaling $\nu=0.005$
(solid line) for the average cross-distance~$d_{12}(t=200)$ (colored bullets)
and initial distances up to $\delta\approx10^{-2}$. 
The gray dots represent the unaveraged 
cross-distances~$d_{12}^{(\mathrm{l})}(t=200)$, as obtained from
distinct $1000$ initial distances. For $\delta>10^{-2}$ the two orbits 
start to end up in distinct attractors, as a second (symmetry related) 
PPC attractor exists close-by in phase space. After a crossover region
$\delta\in[10^{-2},1]$ one observes a second scaling plateau for $\delta>1$.

For the cross-distance scaling of the limit cycle, as
presented for $\rho=\rho_3=181.10$ in Fig.~\ref{fig:breakdown}~($b$), 
we observe an equivalent behavior. In the limit of small initial 
distances~$\delta<10^{-2}$ we obtain in accordance with
Fig.~\ref{fig:distanceScaling} the close to linear scaling $\nu=0.995$.
Again there is a symmetry related limit-cycle attractor close-by in phase space, 
attracting a fraction of the orbits for $\delta>10^{-2}$.

We note that attractors are surrounded, by definition, by a 
possibly small but in any case finite-sized basin of attraction 
\cite{hurley1982attractors} and that the here proposed scaling analysis 
can be performed generically when considering initial conditions close 
enough to the attractor, separated by small initial distances $\delta$. 
Basins of attraction may however fan out 
further away from the attracting set into complicated and possibly 
fractal structures \cite{tel2006chaotic,motter2013doubly}.

\begin{figure*}[t]
\includegraphics[width=.99\textwidth]{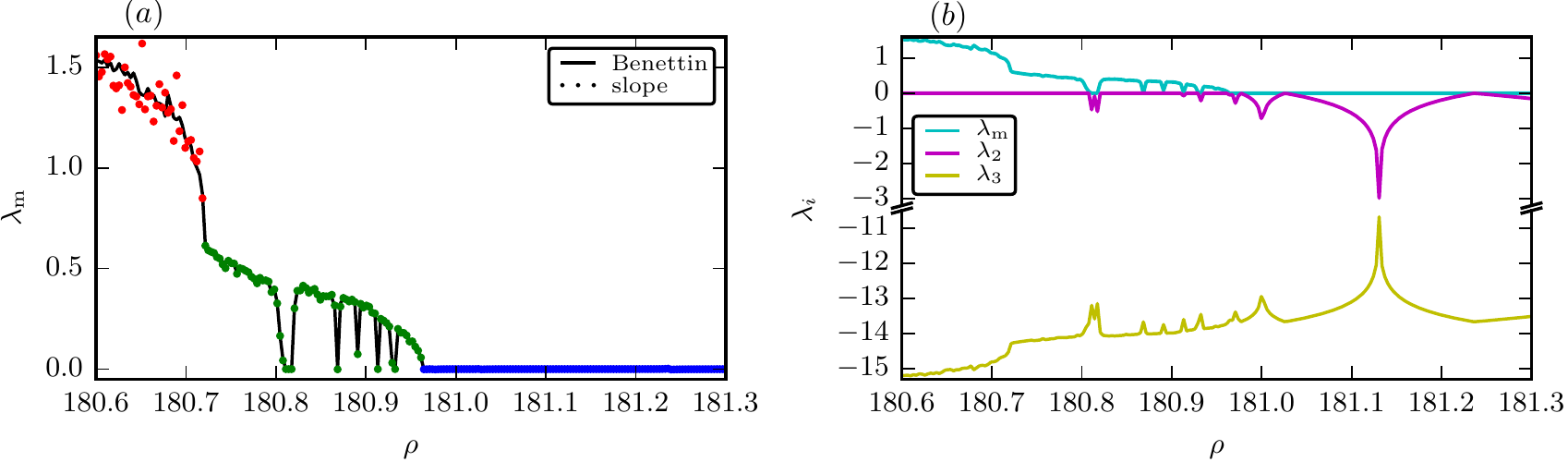}
\caption{\label{fig:lyapcomp} The Lyapunov exponents of the Lorenz 
system for $\rho\in[180.6,181.3]$.
($a$)~Comparing the maximal average Lyapunov exponent $\lambda_\text{m}$ 
obtained using the method of Benettin \cite{benettin1980lyapunov} 
(solid line) and from the initial slope of the 
averaged logarithmic distance 
$\langle\ln\lvert\boldsymbol{x}_1(t)-\boldsymbol{x}_2(t)\rvert\rangle$ (dots).
The distinct regimes are color coded (red/green/blue for strong 
chaos/PPC/laminar flow).
($b$)~All three average Lyapunov exponents 
$\lambda_i$ computed by the method of Benettin 
\cite{benettin1980lyapunov,skokos2010lyapunov}. 
(Note the break in the vertical scale.)}
\end{figure*}

\subsection*{Global Lyapunov exponent}\label{sec:lyapunov}
The results for the maximal average Lyapunov exponent $\lambda_\text{m}$ presented 
above were computed from the averaged logarithmic distance 
$\langle\ln\lvert\boldsymbol{x}_1(t)-\boldsymbol{x}_2(t)\rvert\rangle$ 
between two initially close-by trajectories over time. Alternatively
one may evaluate $\lambda_\text{m}$ using Benettin's 
algorithm\cite{benettin1980lyapunov,skokos2010lyapunov}.

In Fig.~\ref{fig:lyapcomp}~(left) we compare the average maximal
Lyapunov exponent $\lambda_\text{m}$ of the Lorenz system in the 
parameter range $\rho\in[180.6,181.3]$ as obtained by the linear 
slope of the averaged logarithmic distance between two initially 
close-by trajectories (colored dots) with the $\lambda_\text{m}$ 
found when using Benettin's method (solid line). For the latter method 
$\lambda_\text{m}$ is given by the logarithmic ratio of an initial deviation 
from the attractor, 
here $\delta=10^{-8}$, and its stretched time evolution. 
This quantity has been averaged equidistant in time for $\sim10^{6}$ points 
over the respective attractor with an integration time step of $\mathrm 
dt=10^{-3}$. The data matches well for regular motion (blue) and 
for PPC (green). For strong chaos (red) there is however a non-negligible
degree of scattering.

Using Benettin's method we also computed the complete spectrum of 
Lyapunov exponents ($\lambda_\text{m},\lambda_2,\lambda_3$) for the 
Lorenz system in the considered parameter range, as depicted in 
Fig.~\ref{fig:lyapcomp}~(right). The largest and the second largest 
exponents are positive and zero for both chaotic regimes, as expected,
and zero and negative respectively for a limit cycle.
Summing up the exponents leads to 
$\sum_{i}\lambda_i\approx-13.67$, which 
is in agreement with the phase space contraction rate 
$-1-\beta-\sigma=-13.66$ of the Lorenz system.

\begin{figure*}[t]
\includegraphics[width=.99\textwidth]{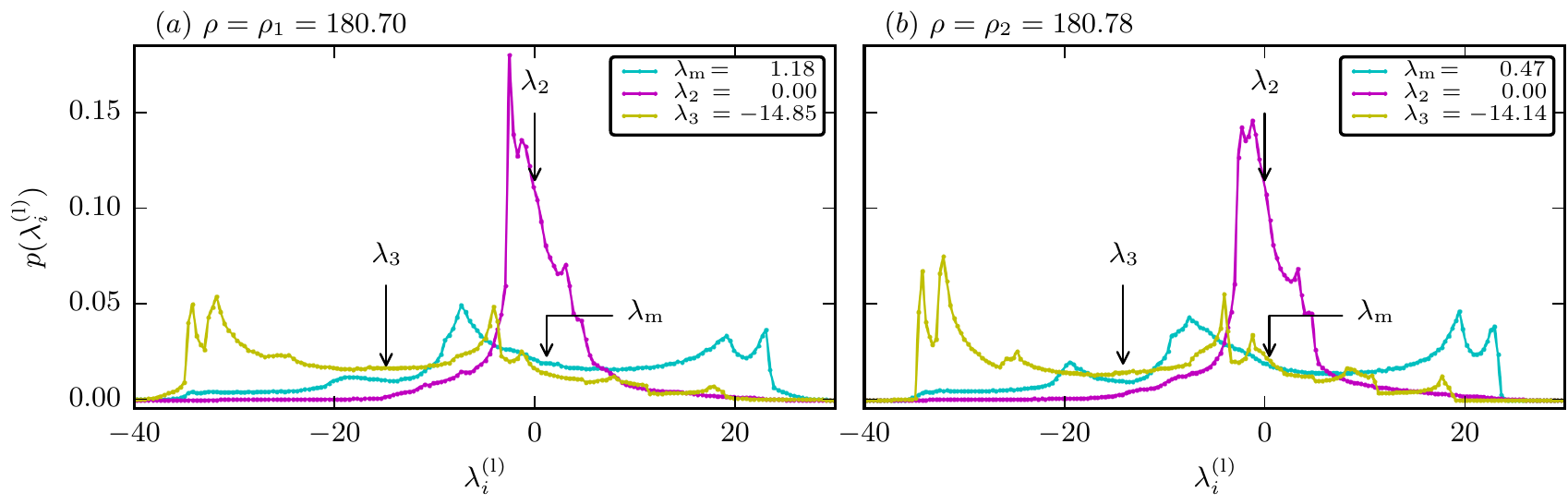}
\caption{\label{fig:lyapspec} The probability distribution $p(\lambda_i^{(\text{l})})$
of the local Lyapunov exponents $\lambda_i^{(\text{l})}$ for the Lorenz attractor;
the global Lyapunov exponents $\lambda_i$ given in the legend
and indicated by arrows in the plot are the average
of the respective distributions. The standard deviations are $s_\text{m}=14.3$, 
$s_2=5.1$ and $s_3=14.3$ in ($a$)~for  $\rho=\rho_1=180.70$ 
(strong chaos) and $s_\text{m}=14.6$, $s_2=5.6$ and
$s_3=14.8$ in ($b$)~for $\rho=\rho_2=180.78$ (PPC). A direct 
connection between the functional form of the distributions of
Lyapunov exponents with the dynamics of PPC is not evident. For
the PPC the variance of the local neutral exponent $\lambda_2^{(l)}$ 
is an order of magnitude larger (5.6 vs.\ 0.47) than the 
corresponding maximal Lyapunov exponent.
}
\end{figure*}

\subsection*{Distribution of local Lyapunov exponents}\label{appsec:locLyap}

It is of interest to evaluate not only the averaged
Lyapunov exponents, as presented in Fig.~\ref{fig:lyapcomp} and
Table~\ref{tab:lyapunov}, but the full distribution of local Lyapunov 
exponents on the attracting set, both for the case of PPC and 
for strong chaos. For the data presented in Fig.~\ref{fig:lyapspec} 
we computed the local Lyapunov exponents $\lambda_i^{(\text{l})}$ 
as the logarithm of the local expansion 
coefficient\cite{skokos2010lyapunov} (the ratio of lengths 
of the orthogonalized deviation vectors after one simulation 
step and the initial deviation vectors), using Gram-Schmidt 
orthogonalization in every step. The local Lyapunov exponents 
were sampled equidistant in time with
an integration time step of $\mathrm dt=10^{-3}$, over a
total time $T_\text{max}=10^{5}$, where the length of initial 
deviation vectors has been set to $\delta=10^{-8}$ after every 
step of the simulation.

The distribution of local Lyapunov exponents covers 
a rather wide range of values as compared to the respective 
global Lyapunov exponents. There is hence no directly 
evident connection between the distribution of local 
Lyapunov exponents and the shape of the chaotic 
attractor, or to the characteristics of PPC. We emphasize that 
the local Lyapunov exponents presented in Fig.~\ref{fig:lyapspec} 
are obtained from the stretching of the deviation vectors 
at every time step (not averaged) and that the
global Lyapunov exponents 
$\lambda_i=\langle\lambda_i^{(\text{l})}\rangle$ 
correspond to the averages of the corresponding local 
exponents over the attracting set.

Fig.~\ref{fig:lyapspec} shows, most interestingly, that
the neutral flow
$\lambda_2=\langle\lambda_2^{(l)}\rangle=0$ is highly
disperse around the chaotic braid. The speed of phase
evolution covers several orders of magnitude. We have 
hence no evidence that the finding that $C_{12}$ remains
diffusive for prolonged time scales, as observed in 
Fig.~\ref{fig:C12_time}~($b$), would result from an effective 
decoupling of a smooth phase and a chaotic radial evolution.

\subsection*{Auto-correlations within PPC}\label{appsec:autocorr}

\begin{figure*}[t]
\includegraphics[width=.99\textwidth]{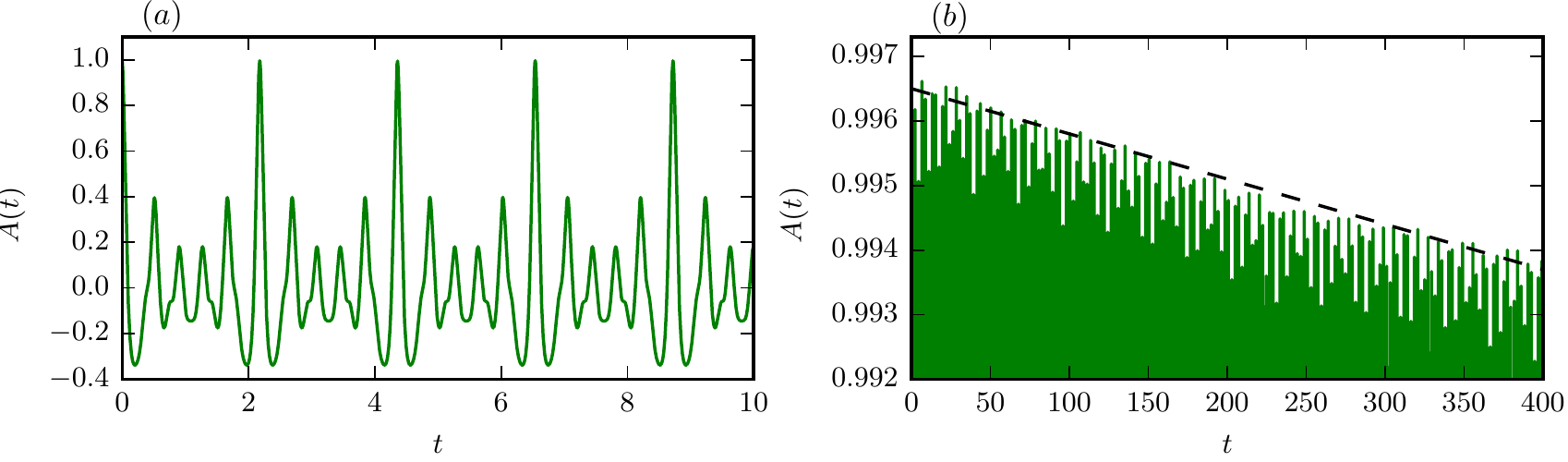}
\caption{\label{fig:autocorr} The autocorrelation $A(t)$, as
defined by Eq.~(\ref{eq:autocorr}), for $\rho=\rho_2=180.78$ (PPC,
compare Fig.~\ref{fig:C12_time}~($b$)).
($a$)~The oscillations resulting from looping around the braids
are spaced by $\tau\simeq2.2$\,.
($b$)~On longer timescales the auto-correlation 
decreases linearly, with the dashed line corresponding to
$0.9965-7\cdot10^{-6}t$.
}
\end{figure*}

Instead of considering the properties of the cross-correlation 
between two trajectories one may study, alternatively, the
autocorrelation function\cite{thomae1981correlations, aizawa1982global,badii1988correlation}
\begin{equation}
A(t)=\lim\limits_{T\to\infty}
\frac{1}{Ts^2}\int_T^{2T}\!\mathrm dt^\prime\;
(\mathbf{x}(t^\prime)-\boldsymbol{\mu})
\,(\mathbf{x}(t^\prime-t)-\boldsymbol{\mu})
\label{eq:autocorr}
\end{equation}
for the trajectory $\mathbf{x}(t)$ on the attractor, where 
$\boldsymbol{\mu}$ and $s$ denote, as defined by Eq.~(\ref{eq:musig}), 
respectively the mean and the average extent of the attracting set.
In Fig.~\ref{fig:autocorr} we present $A(t)$ for the PPC discussed
in Fig.~\ref{fig:C12_time}~($b$), i.\,e.\ for $\rho=\rho_2=180.78$ 
(using $T=10^{4}$). One observes that the topology of the motion 
along fractally broadened braids shows up prominently in the 
autocorrelation function, with the quasi-period $\tau\simeq2.2$ 
of the attractor, determining the separation of the maxima of 
$A(t)$.

The steady loss of predictability observed in 
Fig.~\ref{fig:C12_time}~($b$) for PPC translates into a 
corresponding linear decrease (as indicated by the dashed line
in Fig.~\ref{fig:autocorr}~($b$)) of the heights of the local
maxima of $A(t)$. Using the autocorrelation function for the 
investigation of the slow decorrelation occurring in partially
predictable chaos is hence possible, but plagued by the oscillatory
nature of $A(t)$. For this reason we concentrated in this study
on the cross-correlation $C_{12}(t)$. The initial exponential decrease 
of correlations is furthermore only visible in the data for
$C_{12}(t)$, but not for $A(t)$ (compare
Fig.~\ref{fig:C12_time}~($b$) and Fig.~\ref{fig:autocorr}).


\section*{Acknowledgments}
The authors thank Prof.~Tam\'as T\'el for his suggestions. 
B.\,S.~also acknowledges useful discussions at the 
Seminars in Statistical Physics in Budapest. Further, 
the authors acknowledge the financial support of the 
German research foundation~(DFG) and of Stiftung 
Polytechnische Gesellschaft Frankfurt am Main.

\section*{Author contributions statement}

All authors contributed equally to the present work.
All authors reviewed the manuscript.

\section*{Additional information}

The authors declare no competing financial interests.


\appendix
\section{Appendix A: PPC in a system without separation of scales}\label{appsec:altSys}

\begin{figure*}[t]
\includegraphics[width=.99\textwidth]{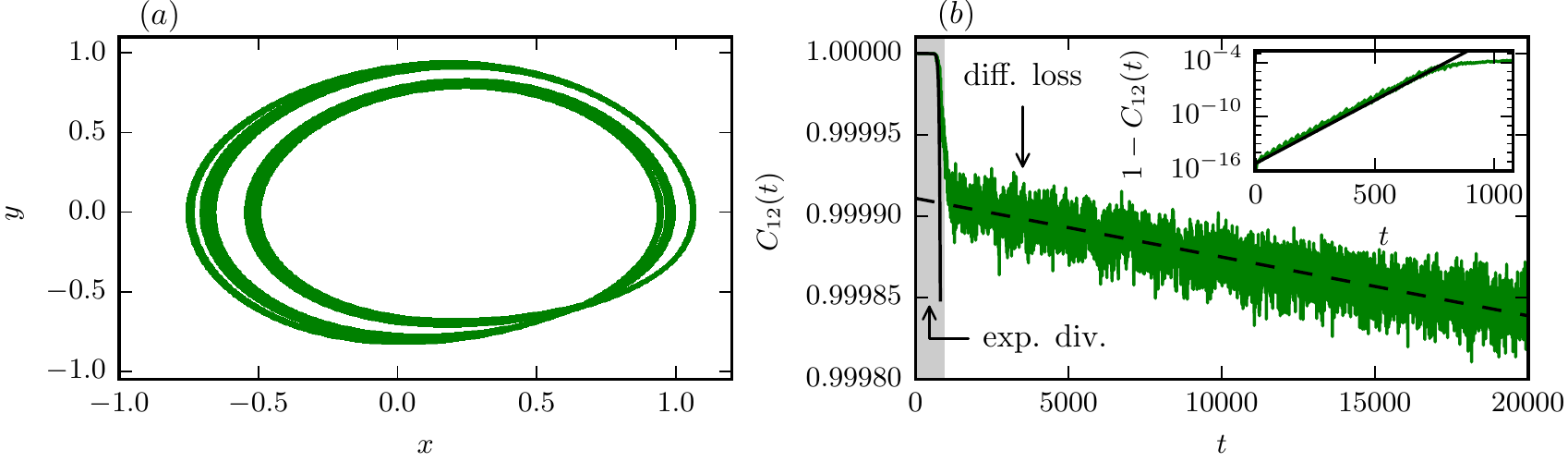}
\caption{\label{fig:shilattr} The attractor of the dynamical system 
described by Eq.~(\ref{eq:shilnikov}), showing partially predictable
chaos for $b=0.3783$.
($a$)~ The projection to the $x-y$ plane showing chaotic braids.
($b$)~The cross-correlation~$C_{12}$, which decreases first
exponentially (gray shaded region), as indicated by the fit $1.3\cdot10^{-9}\e^{\lambda_\text{C}t}$, 
$\lambda_\text{C}=0.032$, (solid line). For larger times the cross-correlation
decreases linearly with a slope $\propto 3.6\cdot10^{-9}t$ (dashed line).}
\end{figure*}

Our choice of the parameters $\beta=8/3$, $\sigma=10$ and 
$\rho>180$ for the Lorenz system (\ref{eq:lorenz}) resulted
in parameters and hence possibly also in time scales of 
distinct orders of magnitude. The question then arises if
the observed large timescale for the final decorrelation process
in the partially predictable phase may be a consequence of
occurrence of distinct microscopic time scales. In order to
rule out this scenario we have investigated with\cite{guan2013non}
\begin{equation}
\label{eq:shilnikov}
\dot{x}=y\,,\quad\qquad
\dot{y}=z\,,\quad\qquad
\dot{z}=x^3-x-y-bz
\end{equation}
a system for which the defining parameters are with 
$b\in[0,1]$ all of the same order of magnitude.
For $b=0.3783$ we find the attractor shown in 
Fig.~\ref{fig:shilattr}~(left), which classifies 
as PPC (partially predictable chaos). The average 
maximal Lyapunov exponent is positive 
($\lambda_\text{m}=0.01, \lambda_2=0,\lambda_3=-0.39$), 
with the chaotic motion being restricted to braids of 
finite width. For slightly larger $b>0.3783$ 
a cascade of period-doubling (halving) bifurcations occurs\cite{guan2013non}.

The cross-correlation shown in Fig.~\ref{fig:shilattr}~(right) 
decays extraordinarily slow (cf.\ Fig.~\ref{fig:C12_time}~($b$)),
with a slope of the order of $10^{-9}$, implying that the retention 
of partial predictability cannot be attributed to the occurrence of 
large intrinsic time scales. The usual initial drop of the 
cross-correlation by about $0.01\%$ within the Lyapunov prediction 
time $T_\lambda\simeq805$ is however present.

\begin{table}[b]\centering
\caption{\label{tab:charac}
Combining the scaling exponent $\nu$ and the level
of the cross-correlation at finite time, $C_{12}(t)$,
allows to classify the three possible types
of dynamics (compare Fig.~\ref{fig:lorenz_overview}).}
\begin{tabular}{ccl}
\toprule
 $\nu$&$C_{12}(t)$&dynamics\\\midrule
 0    & 0         & strong chaos\\
 0    & 1         & PPC\\
 1    & 1         & laminar flow\\
 \bottomrule
\end{tabular}
\end{table}

\section{Appendix B: Automation of the testing procedure}\label{appsec:automat}

As the cross-distance scaling exponent and the finite time cross-correlation
can be used as $0-1$ tests for chaos and PPC respectively, it is possible
to determine whether a system is chaotic, partially predictable or regular,
by an automatized procedure. From the time evolution of pairs of trajectories 
one then determines the maximal Lyapunov exponent $\lambda_\text{m}$, the cross-distance 
scaling exponent $\nu$ and the finite time cross-correlation $C_{12}(t\gg T_\lambda)$, 
where the latter two will be practically binary. The three different dynamics 
can be characterized subsequently by the criteria listed in Table~\ref{tab:charac}.

\begin{enumerate}
\item First one needs to localize the attractor and estimate an upper 
	  bound for the initial distance $\delta\ll1$ (cf.\ the Methods section).
\item One then computes the average maximal Lyapunov exponent 
      $\lambda_\text{m}$ from the slope of the averaged logarithmic distance
      $\langle\log\lvert\mathbf{x}_1(t)-\mathbf{x}_2(t)\rvert\rangle$, where the
      average is performed over pairs of trajectories
      $\mathbf{x}_1(t)$ and $\mathbf{x}_2(t)$ starting from a fixed initial 
      distance $\delta\ll1$. Other methods, like Benettin's 
      algorithm\cite{benettin1980lyapunov}, may be used alternatively.
\item The Lyapunov prediction time $T_\lambda$ is then given by the 
      inverse of the Lyapunov exponent.
\item Next, one computes the average Euclidean distance $d_{12}(t=10\,T_\lambda)$
      for a range of initial distances $\delta\ll1$, from which
       the scaling exponent $\nu$ is extracted using 
       $d_{12}(t>T_\lambda)\sim\delta^\nu$.
       The flow is chaotic for $\nu=0$ and regular for $\nu=1$.
\item For the case of chaotic flows one measures additionally the finite time
      cross-correlation $C_{12}(t=10\,T_\lambda)$ for pairs of trajectories 
      with initial distance $\delta\ll1$.
      The finite time cross-correlation $C_{12}(t=10\,T_\lambda)$ 
      is close to unity and zero respectively for partially predictable
      and for strong chaos.
\end{enumerate}


\end{document}